\documentclass[aps,prb,twocolumn,superscriptaddress,showpacs,amsmath,amssymb]{revtex4}
\pdfoutput=1 
\usepackage{graphicx}
\usepackage{bm}
\usepackage[pdftex,pdfpagemode=UseNone,pdfstartpage=1,pdfstartview=FitH,
	pdfauthor={McFarland, Kott, Sun, Eng, Kane},
	pdftitle={T-dependent transport in a sixfold degenerate 2DES on a H-Si(111) surface}]{hyperref}
\usepackage{color}

\begin{document}
\title{Temperature-dependent transport in a sixfold degenerate two-dimensional electron system on a H-Si(111) surface}

\author{Robert N. McFarland}
\email{robertnm@mailaps.org}
\author{Tomasz M. Kott}
\author{Luyan Sun}
\affiliation{Laboratory for Physical Sciences, University of Maryland
at College Park, College Park, Maryland 20740, USA}%
\author{K. Eng}
\affiliation{Sandia National Laboratories, Albuquerque, New Mexico 87185, USA}
\author{B. E. Kane}
\affiliation{Laboratory for Physical Sciences, University of Maryland
at College Park, College Park, Maryland 20740, USA}%

\date{\today}
\begin{abstract}
Low-field magnetotransport measurements on a high mobility ($\mu$=110,000 cm$^2$/Vs) two-dimensional (2D) electron system on a H-terminated Si(111) surface reveal a sixfold valley degeneracy with a valley splitting $\le$0.1 K.  The zero-field resistivity $\rho_{xx}$ displays strong temperature dependence for 0.07$\le T\le$25 K as predicted for a system with high degeneracy and large mass. We present a method for using the low-field Hall coefficient to probe intervalley momentum transfer (valley drag). The relaxation rate is consistent with Fermi liquid theory, but a small residual drag as $T${$\rightarrow$}0 remains unexplained.
\end{abstract}
\maketitle

Two-dimensional electron systems (2DESs) with additional discrete degrees of freedom (e.g. spin, valleys, subbands, and multiple charge layers) have attracted recent interest due to the role of such variables in transport and in the formation of novel ground states in the the quantized Hall regime. In particular, systems with conduction band valley degeneracy display a rich parameter space for observing and controlling 2DES behavior\cite{ShayeganReview}.  Among multi-valley systems, electrons on the (111) surface of silicon are especially notable because effective mass theory predicts the conduction band to be sixfold degenerate, yielding a total degeneracy (spin$\times$valley) of 12 in the absence of a magnetic field ($B$).  Previous investigations of Si(111) transport using metal-oxide-semiconductor field-effect transistors (MOSFETs) with peak mobilities $\mu\approx4000$ cm$^2$/Vs observed a valley degeneracy $g_v$ of 2 or 6, with the reduced degeneracy attributed to oxide--induced surface strain\cite{ShashkinMass,Tsui6}.  

Here we report transport data on a hydrogen-terminated Si(111) surface (H-Si(111)) with very high mobility ($\mu=110,000$ cm$^2$/Vs at temperature $T$=70 mK and carrier density $n_s=6.7\times10^{11}$ cm$^{-2}$) with clear sixfold valley degeneracy, indicated by the periodicity of Shubnikov-de Haas (SdH) oscillations, isotropic low-$B$ transport, and strong $T$ dependence of the longitudinal resistivity $\rho_{xx}$, consistent with a large $g_v$\cite{Hwang111}.  In addition, we present a method for using the reduced Hall coefficient  $r_H\equiv\rho_{xy}/ (B/en_s)$ in the $B\rightarrow0$ limit as a probe of valley-valley interactions, using a drag model of intervalley momentum transfer in multivalley 2DESs.  We find that the Hall coefficient (and thus, by our model, the intervalley drag) becomes strongly suppressed at low temperatures ($T\lesssim5$K); furthermore, although the $T$ dependence of the drag is roughly quadratic as expected from Fermi liquid theory, a small residual drag in the $T\rightarrow0$ limit remains unexplained.

\begin{figure}
	\begin{center}
	\includegraphics[scale=0.4,draft=false]{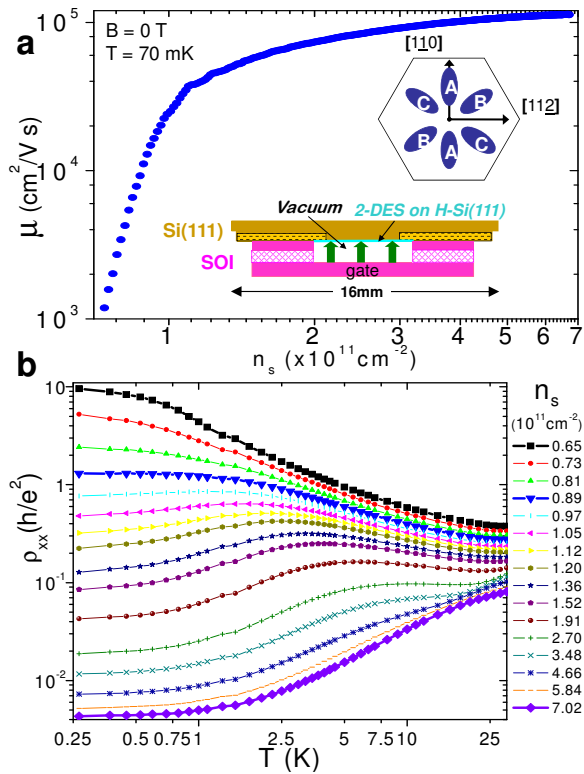}
	\caption[zero field characterization]{\label{mufig}  \textbf{a)} Hall mobility vs.\ carrier density $n_s$. \textit{Upper inset:} Sixfold valley structure of Si(111). \textit{Lower inset:}  Schematic device cross-section showing two of four contacts.  The (500$\mu$m)$^2$ vacuum cavity created by contact bonding preserves the H-{Si(111)} surface and serves as the gate dielectric. \textbf{b)} $\rho$ vs.\ $T$ for $n_s$ ranging from 0.65 ($\blacksquare$ at top) to $7.03\times10^{11}$cm$^{-2}$ (\textcolor[rgb]{.5,0,1}{$\blacklozenge$} on bottom).  The crossover from `metallic' ($d\rho/dT >0$) to `insulating' ($d\rho/dT<0$) behavior occurs near $n_{\textsc{mit}}=0.9\times10^{11}$ cm$^{-2}$(\textcolor{blue}{$\blacktriangledown$}).}
	\end{center}
	\vspace{-18pt}
	\end{figure}

\begin{figure}
	\begin{center}
	\includegraphics[scale=0.3]{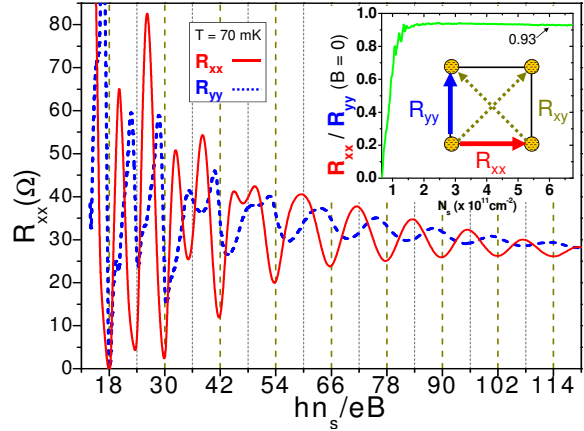}\caption[Low field SdH oscillations]{\label{sdhdualfig} Low-$B$ SdH oscillations for orthogonal current directions.  Below 0.5 T minima appear every 12 levels, indicating a twofold spin and sixfold valley degeneracy. Oscillations continue as high as $\nu\approx234$ ($B\approx0.1$T).  Above 0.5 T further structure becomes evident and minima appear in steps of $\Delta\nu=6$.  While this interval size is essentially the same for both traces, the phase is clearly different. \textbf{Inset:} Base $T$ anisotropy ratio $R_{xx}/R_{yy}$ at $B=0$; diagram shows current configurations for 4-wire measurements of $R_{xx}$, $R_{yy}$, and $R_{xy}$. }
	\end{center}
	\vspace{-18pt}
	\end{figure}

To create and probe a high-mobility electron system on a bare surface, we fabricate a device similar to a four wire MOSFET, with the critical difference that we replace the Si-SiO$_2$ interface with a H-Si(111) surface adjacent to a vacuum cavity\cite{PRL}.  The main processing enhancements in the device discussed here relative to our prior samples are a higher resistivity Si(111) substrate ($\rho\sim10$ k$\Omega$-cm) and final H-termination and bonding performed in an oxygen-free ($<1$ ppm) environment\cite{wet111}.  The resulting device has a very high mobility which, as Fig.\ \ref{mufig} illustrates, increases monotonically with $n_s$ and thus is likely limited by charged impurity scattering due to residual surface charge.  The sample was probed via standard Van der Pauw measurements {(Fig.\ \ref{sdhdualfig} inset)} using low frequency (7-23 Hz) lock-in techniques in both He-3 and dilution refrigerators.

At $B$=0, $\rho_{xx}$ is strongly affected by $T$ (Fig.\ \ref{mufig}b) and at low $T$ displays a metal-insulator crossover near $n_{\textsc{mit}}=0.9\times10^{11}$cm$^{-2}$.  Above this density, the device is clearly metallic, with $\rho$ decreasing by a factor of 3-4 from 5 K to 70 mK at $n_s=7\times10^{11}$cm$^{-2}$.  Compared with Si(100),  Si(111) has a larger density of states effective mass ($m^*_{111}=0.358 m_e$ vs. $m^*_{100}=0.190 m_e$), and a larger $g_v$ (6 vs 2) would lead to a much larger density of states at the Fermi level.  Consequently, electrons on Si(111) should display much stronger screening and therefore a stronger $T$ dependence of $\rho_{xx}$\cite{Hwang111}.  Our observations appear to be qualitatively consistent with such predictions.  At intermediate densities, $\rho_{xx}$ is non-monotonic in $T$, similar to behavior discussed elsewhere \cite{Dilute}.  The data presented in the remainder of this paper was taken at a fixed density of $n_s=6.7\times10^{11}$ cm$^{-2}$.

The low-$B$ SdH oscillations of the present sample reveal minima at filling factors $\nu=hn_s/eB$ spaced by $\Delta\nu=12$ below 0.5 T and $\Delta\nu=6$ above this point (Fig.\ \ref{sdhdualfig}).  Note, however, that the 12-fold-degenerate minima in $\rho_{xx}$ occur at odd multiples of 6 (54, 66, 78, 90...) rather than even as would be expected if the effective mass $m^*$ and $g$-factor $g^*$ are equal to their band values.  This may indicate a $B$-dependent valley splitting and/or enhanced Zeeman splitting at low fields (though unlike the spin-dominated gaps seen in Si(100)\cite{OddNu} and Si(111)\cite{ShashkinMass} MOSFETs, our observations {persist far} in{to} the metallic regime).  At $B>2$ T valley degeneracy lifts, eventually resulting in integer quantum Hall features appearing at intervals of $\Delta\nu=1$ from $\nu=10$ to 1.

Another effect of the weak valley splitting is the observation of isotropic transport ($R_{xx}=R_{yy}$ to within $\approx7\%$) as shown in Fig.\ \ref{sdhdualfig} (inset).  If charge is evenly distributed among a suitably symmetric set of valleys (discussed below), the total resistivity becomes isotropic.  A lifting of the valley degeneracy can shift the valley population into an asymmetric distribution, causing anisotropy in $\rho$ and other transport effects\cite{Gold2Band}.  While previous H-Si(111) devices have shown this anisotropy\cite{PRL}, the absence of such asymmetry in the device presented here suggests that the valley splitting is quite small at $B$=0.   When a small $B$ is applied, the baseline $R_{xx}$ and $R_{yy}$ values remain similar but the {SdH}  response is different: the $\rho_{yy}$ minima  appear out of phase with the  $\rho_{xx}$ minima while retaining the essentially 12-fold periodicity (Fig.\ \ref{sdhdualfig}).  {This may result from a slight population imbalance due to a very small valley splitting, which may increase further for $B>0$.}

In order to provide a quantitative bound on the zero-field valley splitting, we measure the intrinsic (i.e. $B$-independent) level broadening, which is normally characterized in terms of the Dingle temperature $T_D$, or equivalently by the quantum lifetime $\tau_q=\hbar/2\pi k_BT_D$, by examining the evolution of the SdH oscillations as a function of $T$. If the energy level spacing is $E_{gap}=\delta B$ (for some constant $\delta$) the amplitude of these oscillations should be $R_0e^{-2\pi^2 k_BT_D/\delta B}\xi/\sinh(\xi)$ where $\xi=2\pi^2 kT/\delta B$.\cite{AFS,FStilt}  Thus from the $T$ dependence of the amplitudes {(unaffected by the phase anomalies noted above)} we can determine $\tau_q$ as well as the $B$-dependent gap size $\delta$.

\begin{figure}
	\begin{center}
	\includegraphics[scale=0.3]{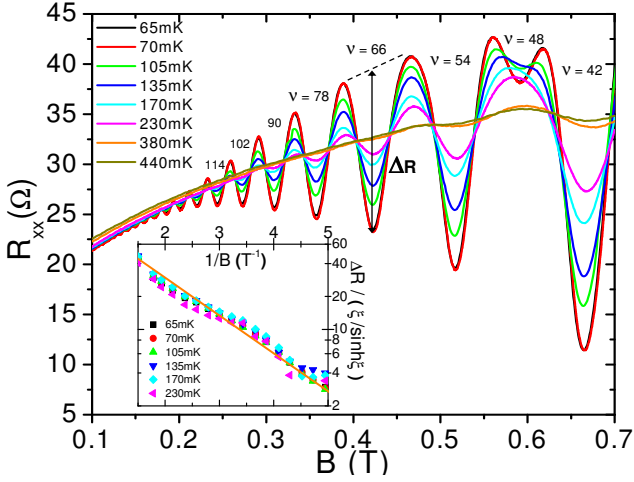}
	\caption[$T$ dependence of SdH oscillations]{\label{sdhTfig}Low-$B$ SdH oscillations at several temperatures. We compute the size $\Delta R$ of the oscillations via linear interpolation of the minima and maxima across a wide range of $\nu$\cite{MassSuppress}. We then plot $\Delta R/T$ vs $T$ to extract the gap size $\delta=E_{gap}/B=2.69$ K/T. \textbf{Inset:} Using this $\delta$, we plot $\Delta R /(\xi/\sinh(\xi))$ vs. $1/B$ to compute the Dingle temperature $T_D=0.1$ K. Note how the data collapse to a single line.}
	\end{center}
	\vspace{-18pt}
	\end{figure}

From the $\rho_{xx}$ oscillations we obtain a value of $\delta=(2.69\pm0.11)$ K/T (Fig.\ \ref{sdhTfig}). Because our gaps occur at odd multiples of six, the simplest analysis would treat them as spin gaps, which would correspond to an enhanced $g^*=4.0\pm0.2$.  However, the anisotropy that emerges when $B>0$ suggests $B$-dependent valley splitting is also present, confounding a simple interpretation of $\delta$.  Regardless of the gaps' origin, from $\delta$ we can compute the quantum lifetime $\tau_q\approx12$ ps, which is quite close to the transport lifetime $\tau_0\approx18$ ps obtained from $\rho_{xx}(B=0)$  as described below; this corresponds to a $T_D$ (and thus an upper bound on the valley splitting) of 0.1~K.

Having established the sixfold valley degeneracy of our 2DES at $B$=0, we now turn to the role of this degeneracy in carrier scattering; in particular we consider the effect of momentum exchange between valleys on low $B$ transport. Semiclassical transport in multiple anisotropic valleys can result in additional (non-oscillatory) $B$ dependence in both $\rho_{xx}$ and $\rho_{xy}$ at low fields ({seen}, for example, in the overall positive slope of the data in Fig.\ \ref{sdhTfig})\cite{PRL}.  This low $B$ behavior can provide information about valley-valley interaction effects.  To see this, we first consider the case of non-interacting valleys that are identical up to rotations $Z(\theta)$ in the $x$-$y$ plane for some $\theta<\pi$  that defines the rotational symmetry of the whole set.  The Drude resistivity for a single valley (with proportional density $n_s/g_v$) for coordinate{s} aligned to the symmetry axes of the valley is given by
	\begin{equation}
	\rho_0=\frac{g_v}{n_se}\left(\begin{array}{ccc}\frac{m_1}{e \tau_0} & & -B \\
	& &\\ B & & \frac{m_2}{e\tau_0} \end{array}\right),
	\end{equation}
where $\tau_0$ is the transport lifetime associated with momentum transfers from the 2DES to the lattice {(both intra- and inter-valley scattering)}.  The resistivity of the $j^{th}$ valley is $\rho_j=Z(j \theta)\rho_0Z(-j \theta)$ and the total {$\rho$} will be
	\begin{equation}
	\label{Eta0}
	\rho=\left(\sum_{j}\rho_j^{-1}\right)^{-1}=\frac{1}{n_se}\frac{1+(\omega_c\tau_0)^2}{\Phi+(\omega_c\tau_0)^2}\left(\begin{array}{ccc}\frac{\bar{m}}{e \tau_0} & & -B \\
	& & \\B & & \frac{\bar{m}}{e \tau_0} \end{array}\right),
	\end{equation}
where $\bar{m}\equiv(m_1+m_2)/2$, $\omega_c=eB/m^*$ is the cyclotron frequency, $m^*=\sqrt{m_1m_2}$, and we define $\Phi\equiv(\bar{m}/m^*)^2$.  For $\omega_c\tau_0\gg1$ both $\rho_{xx}$ and $\rho_{xy}$ are given by their respective classical values $\rho_{xx}=\bar{m}/e^2n_s\tau_0$ and $\rho_{xy}=B/en_s$.  At $B=0$ however, both are suppressed by the factor $\frac{1}{\Phi}\leq1$, with equality only in the case of isotropic valleys ($m_1=m_2$).  In the case of Si(111) (using the band masses  $m_1=0.190 m_e$, $m_2=0.674 m_e$, $m^*=0.358 m_e$\cite{AFS}) we find $\frac{1}{\Phi}=0.686$.

The preceding discussion treats the valleys as independent channels. Strong {valley-valley coupling, as might arise from Coulomb interactions between electrons}, will tend to suppress this correction as {all} electrons move in concert.  To make this {idea} more rigorous, we model intervalley effects as a drag interaction between valleys that conserves total {2DES} momentum while damping the relative momenta between valleys.  {This is distinct from intervalley scattering probed via weak localization\cite{IntervalleyWL} which requires a short-range interaction potential that does not conserve 2DES momentum}.  Following the kinetic approach used in \cite{HwangHall,eeScatter,Hallee} for multi-band systems we obtain a set of coupled equations:

	\begin{equation}
	\frac{M_j \mathbf{v_j}}{\tau_0}  =  e(\mathbf{E}+\mathbf{v_j}\times\mathbf{B})+\frac{1}{\tau_{vv}}\sum_{k\neq j}\left(M_{jk}(\mathbf{v_k}-\mathbf{v_j})\right).
	\end{equation}

Here $M_j$ is the mass tensor of the $j^{th}$ valley, $M_{jk}^{-1}\equiv M_j^{-1}+M_k^{-1}$ is the reduced mass tensor of the j-k system, and $\tau_{vv}$ is the drag relaxation time {(assumed constant and isotropic)}. Combining opposite valleys (which have the same $M_j$), we have three valley pairs.  Substituting $\mathbf{j}_k=n_ke\mathbf{v_k}=n_se\mathbf{v_k}/3$ we then solve the equation $\mathbf{E}=\rho\mathbf{J}=\rho(\mathbf{j_1}+\mathbf{j_2}+\mathbf{j_3})$ for $\rho$.  This gives

	\begin{equation}
	\label{dragxy}
	\rho_{yx}=-\rho_{xy}=\frac{B}{n_se}\frac{(\Lambda\frac{\bar{m}}{m_1}\frac{\tau_0}{\tau_{vv}}+1)(\Lambda\frac{\bar{m}}{m_2}\frac{\tau_0}{\tau_{vv}}+1)+\left(\omega_c\tau_0\right)^2}{\Phi(\Lambda\frac{\tau_0}{\tau_{vv}}+1)^2+\left(\omega_c\tau_0\right)^2}
	\end{equation}
and
	\begin{equation}
	\label{dragxx}
	\rho_{xx}=\rho_{yy}=\frac{\bar{m}}{n_se^2\tau_0}\frac{(\Phi\Lambda\frac{\tau_0}{\tau_{vv}}+1)(\Lambda\frac{\tau_0}{\tau_{vv}}+1)+\left(\omega_c\tau_0\right)^2}{\Phi(\Lambda\frac{\tau_0}{\tau_{vv}}+1)^2+\left(\omega_c\tau_0\right)^2},
	\end{equation}
where $\Lambda\equiv6\det(M_{jk})/\det(M_j)=6/(3\Phi+1)$. In the absence of intervalley interaction, $\frac{\tau_0}{\tau_{vv}}\rightarrow0$ and we recover Eq.\ (\ref{Eta0}).  Conversely, when $\tau_{vv}\ll\tau_0$ we effectively wash out the multi-valley correction.  Thus by measuring $\rho_{xx}$ and $\rho_{xy}$ in the $B=0$ limit we can {solve for} $\tau_0^{-1}$ and $\tau_{vv}^{-1}$.

\begin{figure}
	\begin{center}
	\includegraphics[scale=0.3]{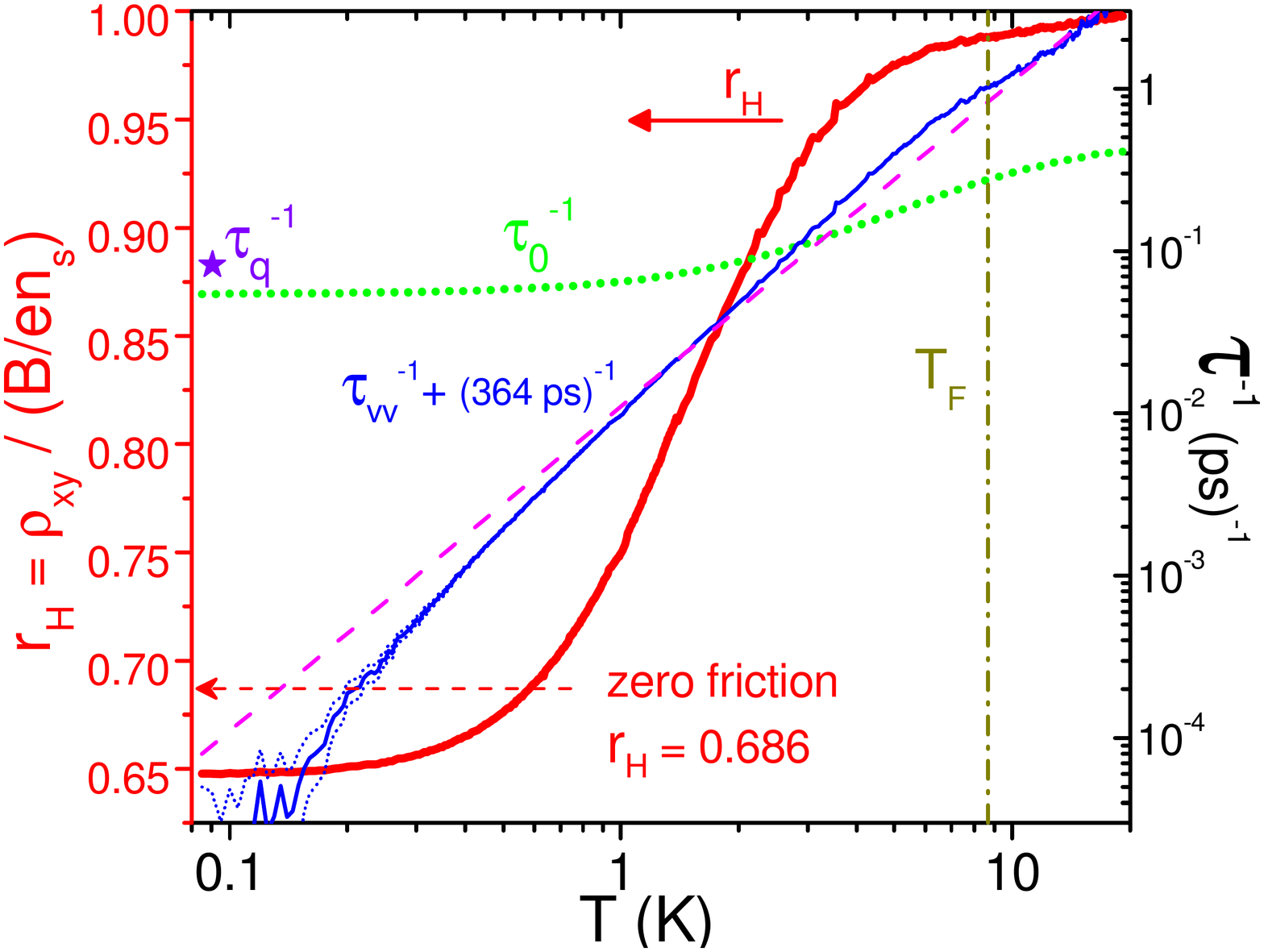}
		\caption[scattering times vs. temperature]{\label{tauvstfig} \textbf{Left axis:} Reduced Hall coefficient $r_H$ measured near $B=0$ (\textcolor{red}{\rule[0.5ex]{1.5em}{0.4ex}}). 	$r_H=0.686$ is the lower bound predicted by the simplest model. \textbf{Right axis:} Lattice scattering rate $\tau_0^{-1}$ (\textcolor{green}{$\centerdot\centerdot\centerdot$}) and valley {drag} relaxation rate $\tau_{vv}^{-1}$ (\textcolor{blue}{\rule[0.5ex]{1.5em}{0.5pt}}) {vs.} $T$, computed from $r_H$ and $\rho_{xx}$ data via Eqs.\ \eqref{dragxy} and \eqref{dragxx}.  The inverse quantum lifetime $\tau_q^{-1}$ (\textcolor[rgb]{.5,0,1}{$\bigstar$}) is calculated from low-$T$ SdH oscillations. The \textcolor[rgb]{1,0,1}{\rule[0.5ex]{1ex}{1pt}\rule{0.5ex}{0pt}\rule[0.5ex]{1ex}{1pt}\rule{0.5ex}{0pt}\rule[0.5ex]{1ex}{1pt}} shows the damping rate expected for a Fermi liquid (see text).}
	\end{center}
	\vspace{-18pt}
	\end{figure}

 Figure \ref{tauvstfig} (left axis) shows such a measurement of $r_H\equiv\rho_{xy}/(B/e n_s)$ vs.\ $T$, averaging results from orthogonal directions (Fig.\ \ref{sdhdualfig}) to remove mixing from $\rho_{xx}$ and $\rho_{yy}$. We determine the density $n_s=6.7\times10^{11}$ cm$^{-2}$ from $R_{xx}$ minima at $\nu=18$ and $\nu=6$ and find it to be insensitive to $T$.  $B$ was held fixed at $\pm50$ mT while $T$ was swept both up and down to ensure consistency.  Taking a slope from these points gives a measure of $r_H$ near $B$=0. Above 5 K, $r_H$ is very close to its classical value, while below 5 K $r_H$ drops rapidly with $T$ before settling to a value of 0.65 at $T=90$ mK.  Interestingly, 0.65 is less than the lower bound of 0.686 predicted {for the drag-free limit}. Because the measurement is based on data taken at $B\neq0$, we expect this to be an overestimate of $r_H$, especially at low $T$ where the $\omega_c\tau_0$ terms in Eqs.\ \eqref{dragxy} and \eqref{dragxx} are largest.  The simplest adjustment we can make to our model to incorporate this discrepancy is to allow $\tau_{vv}^{-1}$ to approach a constant \emph{negative} value at low $T$.

On the right axis of Fig.\ \ref{tauvstfig} we plot the $T$ dependence of the extracted $\tau_{vv}^{-1}$+(364 ps)$^{-1}$  (offsetting $\tau_{vv}^{-1}$ by the base temperature value to remove the divergence) as well as the lattice scattering rate $\tau_0^{-1}$.  The dashed magenta line plots the $T$-dependent electron-electron ($e$-$e$) scattering rate theoretically expected\cite{ZhengScatter} for a single-valley 2D Fermi liquid $\tau_{e\mbox{-}e}^{-1}\sim \frac{E_F}{\hbar}(\frac{T}{T_F})^2$, with a prefactor of 1.9 determined by fitting.  Although earlier work has identified the sensitivity of $r_H$ to $e$-$e$ interactions\cite{HallGaAs,HwangHall,NonMonotonicHall} such corrections are quite small ($\sim$1-4$\%$) at high densities ($n_s\gg n_{\textsc{mit}}$).  Furthermore, these models predict $r_H\rightarrow1$ in the $T\rightarrow0$ limit, whereas our model treats the multi-valley effects of Eq.\ (\ref{Eta0}) as intrinsic at $T=0$.

Several open questions remain regarding this data.  First is {the dominance of the odd gaps and the low-field phase anisotropy in the SdH data.  Although the experimental setup did not allow for tilted B field measurements, such experiments could help identify the roles of cyclotron, spin, and valley splitting in forming these gaps.  Second, we note} the disparity between the measured $r_H$ at base temperature and the lower bound given by our model, which we have presented in terms of negative drag for mathematical simplicity. Known corrections to $\rho_{xy}$, for instance due to disorder, would \emph{increase} $r_H$\cite{HallGaAs,HwangHall,NonMonotonicHall}, not decrease it as reported here.  Negative drag has been discussed as a possible consequence of electron correlation in bilayer systems\cite{SignReversal,DragExtreme,NegDrag}. Small sample anisotropies (neglected in our model) could also play a role.   Another possibility is that anisotropic enhancement of $m^*$ could change  the $T=0$ limit of Eq.\ (\ref{dragxy}) (note that \emph{isotropic} enhancement would not change $r_H$, which depends only on mass \emph{ratios}).  Alternatively, a more sophisticated theory of valley-valley interactions could modify the $T\rightarrow0$ limit of our simple model.  

Finally, we consider why sixfold degeneracy persists in this device given the ease with which mechanisms such as misorientation, disorder, and strain can lift this degeneracy\cite{AFS,Tsui6,MiscutTight} and that previous work on H-Si(111) found the {$g_v=6$} ground state split into a low energy {$g_v=2$} band with a {$g_v=4$} band $\sim7$ K above it\cite{PRL}.  If that gap was in fact produced or enhanced by surface disorder, perhaps the higher mobility of the present sample can account for the difference.  Further work on the relationships between surface preparation, device mobility, and valley splitting is presently underway.

This work was funded by the Laboratory for Physical Sciences.  Sandia National Laboratories is a multi-program laboratory operated by Sandia Corporation, a Lockheed-Martin Co., for the U. S. Department of Energy under Contract No. DE-AC04-94AL85000.


\begin{thebibliography}{26}

\providecommand{\bibinfo}[2]{#2}

\bibitem[{{Shayegan et~al.}(2006){Shayegan, Poortere,
  Gunawan, Shkolnikov, Tutuc, and Vakili}}]{ShayeganReview}
\bibinfo{author}{{M.}~{Shayegan}},
  \bibinfo{author}{{E.~P.~D.} {Poortere}},
  \bibinfo{author}{{O.}~{Gunawan}},
  \bibinfo{author}{{Y.~P.} {Shkolnikov}},
  \bibinfo{author}{{E.}~{Tutuc}}, {and}
  \bibinfo{author}{{K.}~{Vakili}},
  \bibinfo{journal}{Phys.\ Status Solidi B} \textbf{\bibinfo{volume}{243}},
  \bibinfo{pages}{3629} (\bibinfo{year}{2006}).

\bibitem[{{{Shashkin} et~al.}(2007){{Shashkin},
  {Kapustin}, {Deviatov}, {Dolgopolov}, and {Kvon}}}]{ShashkinMass}
\bibinfo{author}{{A.~A.} {{Shashkin}}},
  \bibinfo{author}{{A.~A.} {{Kapustin}}},
  \bibinfo{author}{{E.~V.} {{Deviatov}}},
  \bibinfo{author}{{V.~T.} {{Dolgopolov}}},
  {and} \bibinfo{author}{{Z.~D.}
  {{Kvon}}}, \bibinfo{journal}{\prb} \textbf{\bibinfo{volume}{76}},
  \bibinfo{pages}{241302(R)} (\bibinfo{year}{2007}).

\bibitem[{{Tsui and Kaminsky}(1979)}]{Tsui6}
\bibinfo{author}{{D.~C.} {Tsui}} {and}
  \bibinfo{author}{{G.}~{Kaminsky}},
  \bibinfo{journal}{Phys.\ Rev.\ Lett.} \textbf{\bibinfo{volume}{42}},
  \bibinfo{pages}{595} (\bibinfo{year}{1979}).

\bibitem[{{Hwang and {Das Sarma}}(2007)}]{Hwang111}
\bibinfo{author}{{E.~H.} {Hwang}} {and}
  \bibinfo{author}{{S.}~{{Das Sarma}}},
  \bibinfo{journal}{\prb} \textbf{\bibinfo{volume}{75}}, \bibinfo{eid}{073301}
  (\bibinfo{year}{2007}).

\bibitem[{{{Eng} et~al.}(2007){{Eng}, {McFarland},
  and {Kane}}}]{PRL}
\bibinfo{author}{{K.}~{{Eng}}},
  \bibinfo{author}{{R.~N.} {{McFarland}}},
  {and} \bibinfo{author}{{B.~E.}
  {{Kane}}}, \bibinfo{journal}{\prl} \textbf{\bibinfo{volume}{99}},
  \bibinfo{pages}{016801} (\bibinfo{year}{2007}).  
   \bibinfo{journal}{\apl} \textbf{\bibinfo{volume}{87}}, \bibinfo{eid}{052106}
  (\bibinfo{year}{2005}).

\bibitem[{{Nishizawa et~al.}(2006){Nishizawa,
  Bolotov, Tada, and Kanayama}}]{wet111}
\bibinfo{author}{{M.}~{Nishizawa}},
  \bibinfo{author}{{L.}~{Bolotov}},
  \bibinfo{author}{{T.}~{Tada}}, {and}
  \bibinfo{author}{{T.}~{Kanayama}},
  \bibinfo{journal}{J. Vac. Sci. Technol. B} \textbf{\bibinfo{volume}{24}},
  \bibinfo{pages}{365} (\bibinfo{year}{2006}).

\bibitem[{{Punnoose and Finkel'stein}(2001)}]{Dilute}
\bibinfo{author}{{A.}~{Punnoose}} {and}
  \bibinfo{author}{{A.~M.} {Finkel'stein}},
  \bibinfo{journal}{Phys.\ Rev.\ Lett.\ } \textbf{\bibinfo{volume}{88}},
  \bibinfo{pages}{016802} (\bibinfo{year}{2001}).

\bibitem[{{Kravchenko et~al.}(2000){Kravchenko,
  Shashkin, Bloore, and Klapwijk}}]{OddNu}
\bibinfo{author}{{S.~V.} {Kravchenko}},
  \bibinfo{author}{{A.~A.} {Shashkin}},
  \bibinfo{author}{{D.~A.} {Bloore}},
  {and} \bibinfo{author}{{T.~M.}
  {Klapwijk}}, \bibinfo{journal}{Solid State Comm.}
  \textbf{\bibinfo{volume}{116}}, \bibinfo{pages}{495 } (\bibinfo{year}{2000}).

\bibitem[{{{Gold} et~al.}(2008){{Gold}, {Fabie}, and
  {Dolgopolov}}}]{Gold2Band}
\bibinfo{author}{{A.}~{{Gold}}},
  \bibinfo{author}{{L.}~{{Fabie}}}, {and}
  \bibinfo{author}{{V.~T.} {{Dolgopolov}}},
  \bibinfo{journal}{Physica E} \textbf{\bibinfo{volume}{40}},
  \bibinfo{pages}{1351} (\bibinfo{year}{2008}).

\bibitem[{{Ando et~al.}(1982){Ando, Fowler, and
  Stern}}]{AFS}
\bibinfo{author}{{T.}~{Ando}},
  \bibinfo{author}{{A.~B.} {Fowler}},
  {and} \bibinfo{author}{{F.}~{Stern}},
  \bibinfo{journal}{Rev. Mod. Phys.} \textbf{\bibinfo{volume}{54}},
  \bibinfo{pages}{437} (\bibinfo{year}{1982}).

\bibitem[{{Fang and Stiles}(1968)}]{FStilt}
\bibinfo{author}{{F.~F.} {Fang}} {and}
  \bibinfo{author}{{P.~J.} {Stiles}},
  \bibinfo{journal}{Phys. Rev.} \textbf{\bibinfo{volume}{174}},
  \bibinfo{pages}{823} (\bibinfo{year}{1968}).

\bibitem[{{Padmanabhan et~al.}(2008){Padmanabhan,
  Gokmen, Bishop, and Shayegan}}]{MassSuppress}
\bibinfo{author}{{M.}~{Padmanabhan}},
  \bibinfo{author}{{T.}~{Gokmen}},
  \bibinfo{author}{{N.~C.} {Bishop}},
  {and} \bibinfo{author}{{M.}~{Shayegan}},
  \bibinfo{journal}{\prl} \textbf{\bibinfo{volume}{101}},
  \bibinfo{pages}{026402} (\bibinfo{year}{2008}).

\bibitem[{{Kuntsevich et~al.}(2007){Kuntsevich,
  Klimov, Tarasenko, Averkiev, Pudalov, Kojima, and
  Gershenson}}]{IntervalleyWL}
\bibinfo{author}{{A.~Y.} {Kuntsevich}},
\bibinfo{author}{{N.~N.} {Klimov}},
  \bibinfo{author}{{S.~A.} {Tarasenko}},
  \bibinfo{author}{{N.~S.} {Averkiev}},
  \bibinfo{author}{{V.~M.} {Pudalov}},
  \bibinfo{author}{{H.}~{Kojima}}, {and}
  \bibinfo{author}{{M.~E.} {Gershenson}},
  \bibinfo{journal}{\prb} \textbf{\bibinfo{volume}{75}},
  \bibinfo{pages}{195330} (\bibinfo{year}{2007}).

\bibitem[{{Hwang and {Das Sarma}}(2006)}]{HwangHall}
\bibinfo{author}{{E.~H.} {Hwang}} {and}
  \bibinfo{author}{{S.}~{{Das Sarma}}},
  \bibinfo{journal}{Phys. Rev. B} \textbf{\bibinfo{volume}{73}},
  \bibinfo{pages}{121309(R)} (\bibinfo{year}{2006}).

\bibitem[{{Kukkonen and Maldague}(1979)}]{eeScatter}
\bibinfo{author}{{C.~A.} {Kukkonen}} {and}
  \bibinfo{author}{{P.~F.} {Maldague}},
  \bibinfo{journal}{Phys.\ Rev.\ B} \textbf{\bibinfo{volume}{19}},
  \bibinfo{pages}{2394} (\bibinfo{year}{1979}).

\bibitem[{{Vitkalov}(2001)}]{Hallee}
\bibinfo{author}{{S.~A.} {Vitkalov}},
  \bibinfo{journal}{\prb} \textbf{\bibinfo{volume}{64}},
  \bibinfo{pages}{195336} (\bibinfo{year}{2001}).

\bibitem[{{Zheng and {Das Sarma}}(1996)}]{ZhengScatter}
\bibinfo{author}{{L.}~{Zheng}} {and}
  \bibinfo{author}{{S.}~{{Das Sarma}}},
  \bibinfo{journal}{Phys.\ Rev.\ B} \textbf{\bibinfo{volume}{53}},
  \bibinfo{pages}{9964} (\bibinfo{year}{1996}).

\bibitem[{{Yasin et~al.}(2005){Yasin, Sobey,
  Micolich, Clarke, Hamilton, Simmons, Pfeiffer, West, Linfield, Pepper, Richie.}}]{HallGaAs}
\bibinfo{author}{{C.~E.} {Yasin}},
\bibinfo{author}{{T.~L.} {Sobey}},
  \bibinfo{author}{{A.~P.} {Micolich}},
  \bibinfo{author}{{W.~R.} {Clarke}},
  \bibinfo{author}{{A.~R.} {Hamilton}},
  \bibinfo{author}{{M.~Y.} {Simmons}},
  \bibinfo{author}{{L.~N.} {Pfeiffer}},
  \bibinfo{author}{{K.~W.} {West}},
  \bibinfo{author}{{E.~H.} {Linfield}},
  \bibinfo{author}{{M.}~{Pepper}},
  \bibinfo{author}{{D.~A.}~{Richie}}, \bibinfo{journal}{Phys. Rev. B}
  \textbf{\bibinfo{volume}{72}}, \bibinfo{pages}{241310(R)}
  (\bibinfo{year}{2005}).

\bibitem[{{Kuntsevich et~al.}(2005){Kuntsevich,
  Knyazev, Kozub, Pudalov, Brunhaler, and Bauer}}]{NonMonotonicHall}
\bibinfo{author}{{A.~Y.} {Kuntsevich}},
\bibinfo{author}{{D.~A.} {Knyazev}},
  \bibinfo{author}{{V.~I.} {Kozub}},
  \bibinfo{author}{{V.~M.} {Pudalov}},
  \bibinfo{author}{{G.}~{Brunhaler}},
  {and} \bibinfo{author}{{G.}~{Bauer}},
  \bibinfo{journal}{JETP Lett.} \textbf{\bibinfo{volume}{81}},
  \bibinfo{pages}{409} (\bibinfo{year}{2005}).

\bibitem[{{Alkauskas et~al.}(2002){Alkauskas,
  Flensberg, Hu, and Jauho}}]{SignReversal}
\bibinfo{author}{{A.}~{Alkauskas}},
  \bibinfo{author}{{K.}~{Flensberg}},
  \bibinfo{author}{{Ben~Yu-Kuang}~{Hu}}, 
  {and}  \bibinfo{author}{{A.-P.} {Jauho}},
  \bibinfo{journal}{Phys.\ Rev.\ B} \textbf{\bibinfo{volume}{66}},
  \bibinfo{pages}{201304(R)} (\bibinfo{year}{2002}).

\bibitem[{{Lilly et~al.}(1998){Lilly, Eisenstein,
  Pfeiffer, and West}}]{DragExtreme}
\bibinfo{author}{{M.~P.} {Lilly}},
  \bibinfo{author}{{J.~P.} {Eisenstein}},
  \bibinfo{author}{{L.~N.} {Pfeiffer}},
  {and} \bibinfo{author}{{K.~W.} {West}},
  \bibinfo{journal}{Phys.\ Rev.\ Lett.\ } \textbf{\bibinfo{volume}{80}},
  \bibinfo{pages}{1714} (\bibinfo{year}{1998}).

\bibitem[{{Price et~al.}(2007){Price, Savchenko,
  Narozhny, Allison, and Ritchie}}]{NegDrag}
\bibinfo{author}{{A.~S.} {Price}},
  \bibinfo{author}{{A.~K.} {Savchenko}},
  \bibinfo{author}{{B.~N.} {Narozhny}},
  \bibinfo{author}{{G.}~{Allison}}, {and}
  \bibinfo{author}{{D.~A.} {Ritchie}},
  \bibinfo{journal}{Science} \textbf{\bibinfo{volume}{316}},
  \bibinfo{pages}{99} (\bibinfo{year}{2007}).

\bibitem[{{Kharche et~al.}(2008){Kharche, Kim,
  Boykin, and Klimeck}}]{MiscutTight}
\bibinfo{author}{{N.}~{Kharche}},
  \bibinfo{author}{{S.}~{Kim}},
  \bibinfo{author}{{T.~B.} {Boykin}},
  {and} \bibinfo{author}{{G.}~{Klimeck}},
  \bibinfo{journal}{Appl.\ Phys.\ Lett.} \textbf{\bibinfo{volume}{94}},
  \bibinfo{pages}{042101} (\bibinfo{year}{2009}).

\end{thebibliography}
\end{document}